\documentclass[aps,pre,twocolumn,superscriptaddress,groupedaddress]{revtex4}  
\usepackage{amsfonts}
\usepackage{amsmath}
\usepackage{amssymb}
\usepackage{version}
\usepackage{graphicx}%
\includeversion{New_connection}
\excludeversion{Old_connection}

\begin{document}
\title{Switching Current Distributions in Josephson Junctions at Very Low Temperatures}
\author{James A. Blackburn}
\affiliation{Physics \& Computer Science, Wilfrid Laurier University, Waterloo, Ontario, Canada}
\author{Matteo Cirillo}
\affiliation{Dipartimento di Fisica and MINAS-Lab, Universit\`{a} di Roma \textquotedblleft Tor Vergata\textquotedblright\ I-00133 Roma, Italy}
\affiliation{\textit{CNR-SPIN} Institute, Italy}
\author{Niels Gr{\o}nbech-Jensen}
\affiliation{Department of Applied Science, University of California, Davis, California 95616}

\begin{abstract}
Swept bias experiments carried out on Josephson junctions yield the
distributions of the probabilities of early switching from the zero voltage
state. \ Kramers' theory of thermally activated escape from a one-dimensional
potential is well known to fall short of explaining such experiments when the
junctions are at millikelvin temperatures. We propose a simple revision of the
theory which is shown to yield extremely good agreement with experimental data.
\end{abstract}
\maketitle

When a Josephson junction is biased with a steady current the phase variable
behaves like the position coordinate that describes a \textquotedblleft
particle\textquotedblright\ sitting at the bottom of a well in a washboard
potential \cite{Tinkham}. An external dc bias current has the effect of
tilting the washboard and causing the wells to become more shallow and
disappear when the bias is equal to the junction critical current. \ At bias
currents smaller than the critical value, the \textquotedblleft
particle\textquotedblright\ remains in a well unless it is able to escape from
it via some mechanism. The first escape mechanism to be recognized was
classical thermal activation (TA) \cite{Kramers} in which the
\textquotedblleft particle\textquotedblright\ jumps over the barrier and then
bounces down the washboard. \ In Josephson terms this effect results in a
non-zero dc voltage across the junction.

The analogy of the Josephson potential with the general class of
one-dimensional (single coordinate) potentials makes the investigation of its
features particularly interesting. One of the most significant developments in
recent years began with the experiments of Voss \& Webb \cite{Voss}. A
Josephson junction was repeatedly subjected to a smoothly increasing bias. The
distribution of values of the current at which the junction switched to a
finite voltage state, also termed the switching current distribution (SCD),
was recorded. The beauty of the original swept bias experiment was its
conceptual simplicity. In a previous paper \cite{Blackburn2012} we noted that
the escape peaks reported in \cite{Voss} did not freeze at a given temperature
but rather continued to advance towards higher bias currents as the
temperature was lowered. We raised the possibility that this behavior could be
evidence of classical escape dynamics.

In the present paper, using more recent experimental data by Yu {\it et al.}
\cite{Yu}, we assess in greater detail the predictions of TA theory. This
source of data was chosen because that experiment was fully characterised and,
most particularly, because in Fig.~2 of \cite{Yu} data pairs are shown for the
observed peak positions and widths at 19 different temperatures ranging from
just below $0.8K$ down to $25mK$. \ Observations very similar to those in
\cite{Yu} have been reported many times in the literature, e.g.,\cite{Li, Cui,
Longobardi, Yu2}, which leads us to be confident that our conclusions are
\textit{generally} applicable to Josephson junctions at low temperatures. We
draw the following two conclusions from the work reported here: first, that a
small elevation of the sample temperature above the dilution refrigerator
temperature is sufficient to explain experimental observations using a
classical thermal activation model, and second, that published data do not
exhibit saturation at temperatures below a presumptive crossover as required
by MQT theory.

A full description of the simulation model for a swept bias experiment was
presented in \cite{Blackburn2012}. The key input parameters for the computer
program are: the Josephson critical current (for Yu {\it et al.} $1.957\mu A$), the
plasma frequency (for Yu {\it et al.} $15.59GHz$), the bias ramp time (for Yu {\it et al.}
$4.89m\sec$), and the temperature.

A central role is played by the escape rate $\Gamma(t)$ which for TA theory
applied to a Josephson junction is%
\begin{equation}
\Gamma(t)=f\exp\left(  -\frac{\Delta U}{k_{B}T_{J}}\right)  \label{Kramers}%
\end{equation}
where $f$ is the plasma frequency for the well, $\Delta U$ \ is the height of
the potential barrier, and $T_{J}$ is the junction temperature.

\begin{figure}[t]
\begin{center}
\scalebox{0.3}{\centering
\includegraphics{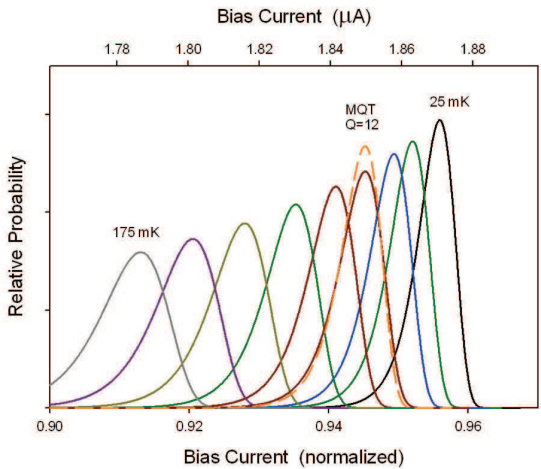}}
\end{center}
\vspace{-0.5cm}
\caption{Escape probability distributions (simulation) at various
temperatures. The MQT distribution (dashed line) was simulated with $Q=12$ and
coincides with the thermal activation peak at $T=65mK$.}
\label{Fig1}
\end{figure}

Using the system values given in \cite{Yu}, the swept bias simulation was run
at various junction temperatures.\ The results are shown in Fig.~\ref{Fig1}.%

These simulation peaks closely match the experimental data represented in
Fig.~1 of \cite{Yu}.

\begin{figure}[t]
\begin{center}
\scalebox{0.35}{\centering
\includegraphics
{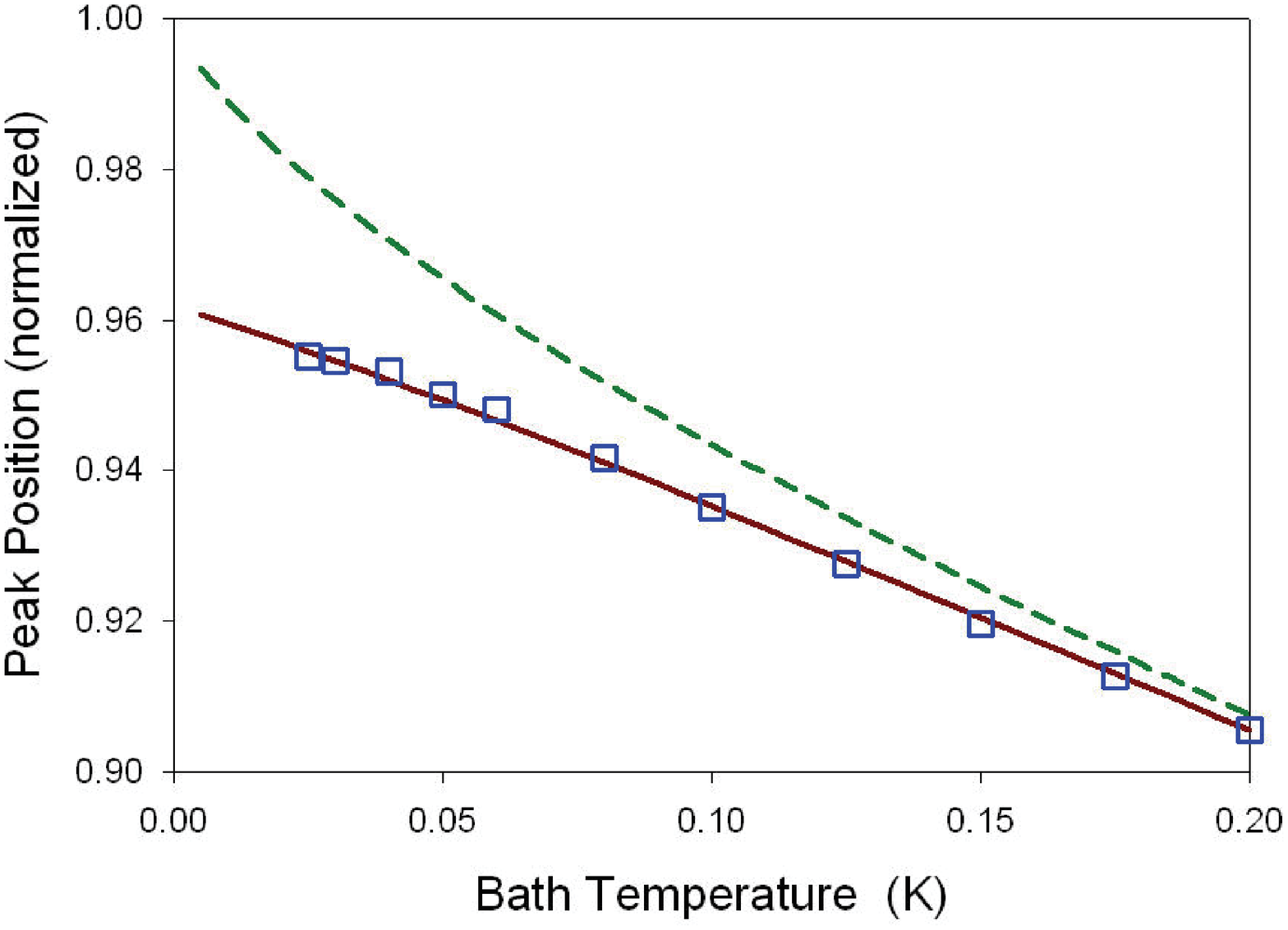}}%
\caption{Experimentally observed escape probability peak positions \cite{Yu}
as a function of bath temperature (blue squares). \ The dashed line is the
classical thermal activation result Eq.~(\ref{Kramers}). \ The solid line is
the result of the modified classical theory Eq.~(\ref{Kramers2}).}%
\label{Fig2}%
\end{center}
\end{figure}

We now consider the behavior of the positions of the peaks of Fig.~1 as a
function of temperature. In Yu {\it et al.} \cite{Yu} the experimental data for peak
positions are shown as squares in their Fig.~2. \ We plot their $11$ lowest
temperature points in Fig.\ref{Fig2} together with our simulation results
based on the classical thermal escape rate expressed by Eq.~(\ref{Kramers}).
\ Note that we use a linear temperature axis in this plot.%

The experimental data for peak positions fall below the TA theory results at
low temperatures. This separation of theory from experiment is \textit{smooth}%
, suggesting a continuous process rather than a definite transition to any
different underlying escape mechanism. The TA simulation outcome (dashed
curve) can be made to drop\ to the experimental data points (squares) by
modifying the classical escape rate expression (\ref{Kramers}) as follows%
\begin{equation}
\Gamma(t)=f\exp\left(  -\varepsilon\frac{\Delta U}{k_{B}T}\right)
\label{Kramers2}%
\end{equation}
It was found that%
\begin{equation}
\varepsilon=1.0-\exp(-gT) \label{epsilon}%
\end{equation}
with $g=17.5K^{-1}$ yielded a revised classical result (solid curve in
Fig.~\ref{Fig2}) that is in outstanding agreement with the experimental data.

\begin{figure}[t]
\begin{center}
\scalebox{0.3}{\centering
\includegraphics
{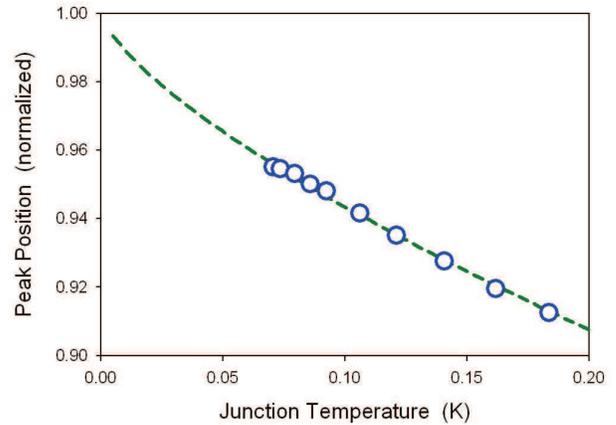}}%
\caption{Dependence of escape peak positions as a function of junction
temperature from a simulation based of the escape rate Eq.~(\ref{Kramers}).
\ Experimental data from \cite{Yu} (circles) have been converted from bath
temperature to junction temperature.}%
\label{Fig3}%
\end{center}
\end{figure}

The exponent in Eq.~(\ref{Kramers2}) can be regarded as containing a modified
junction temperature $T\rightarrow T/\varepsilon$, with larger modifications
occurring at low values of $T$. \ Therefore $T$ should be identified as the
experimentally known bath temperature $T_{B}$. The key point is that $T_{B}$
is provided by the cryogenic apparatus whereas the all-important junction
temperature $T_{J}$ is not known. In a situation where $T_{J}\neq T_{B}$,
using $T_{B}$ in place of $T_{J}$ in Eq.~(\ref{Kramers}) will make it
\textit{appear} that the classical theory has somehow failed. It should be
noted that from this perspective, identifying, as is commonly done, the bath
temperature with the junction temperature when plotting experimental data for
peak position versus temperature is fundamentally wrong, most especially at
low bath temperatures where the junction temperature might be elevated. To
illustrate this, we use the expression $T_{J}=T_{B}/\varepsilon$ to `correct'
the temperatures of experimental data points derived from Fig.~2 in \cite{Yu}.
\ With these junction temperatures in place of the bath temperatures, we
obtain the results shown in Fig.~\ref{Fig3}.%
Comparing this figure with Fig.~\ref{Fig2}, it is now clear that the classical
TA theory gives an excellent accounting of experiment - if the experimental
bath temperatures are understood to be not the same as the temperatures of the junction.

\begin{figure}[t]
\begin{center}
\scalebox{0.3}{\centering
\includegraphics
{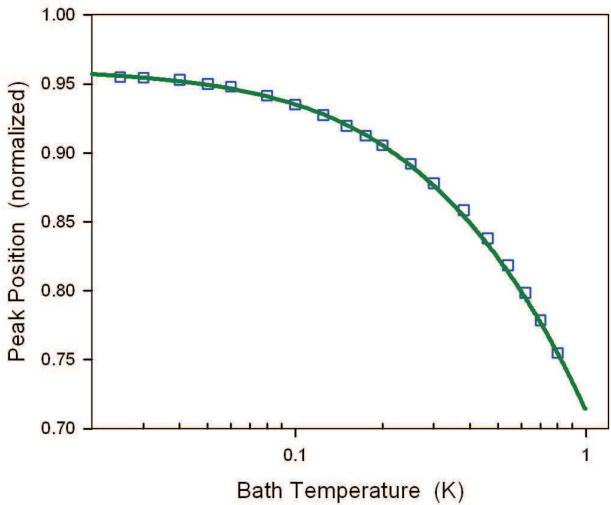}}%
\caption{Peak position versus temperature. Squares are experimental data
points from \cite{Yu}. The solid line was produced from the modified classical
escape rate Eq.~(\ref{Kramers2}).}%
\label{Fig4}%
\end{center}
\end{figure}

Figure \ref{Fig4} covers the temperature range reported in \cite{Yu} and
adopts the same logarithmic temperature scale. Here the success of the
modified exponent in matching theory and experiment over the full temperature
range is quite apparent.%

The same question regarding the apparent discrepancy between experiment and TA
theory was answered differently in Voss \& Webb \cite{Voss}. They invoked the
hypothesis that below a crossover temperature the junction phase would behave
as a macroscopic quantum variable and the junction could escape to a running
state by tunneling out of the well. In the MQT regime the escape rate is
expected to be \cite{Devoret}%

\begin{equation}
\Gamma(t)=a_{q}f\exp\left[  -7.2\frac{\Delta U}{hf}\left[  1+\frac{0.87}%
{Q}\right]  \right]  \label{MQT}%
\end{equation}
where
\begin{equation}
a_{q}=\left[  120\pi\left(  \frac{7.2\Delta U}{hf}\right)  \right]  ^{%
\frac12
} \label{aq}%
\end{equation}

We performed a bias sweep simulation using the MQT escape rate given in the
above Eq.~ (\ref{MQT}) setting $Q=12$ \cite{Q}$.$ The resulting SCD peak is
included in Fig.~\ref{Fig1} where it can be seen to coincide with the SCD peak
for TA at $T=65mK$ - essentially the crossover temperature. Note that for TA
the escape rate is temperature dependent whereas the MQT escape rate
Eq.~(\ref{MQT}) does not include temperature.\ A direct consequence of this is
that following a transition into the macroscopic quantum regime, if that
occurs, the resulting escape process should become temperature independent.
However such a change is not what the experimental data exhibit -- the
experimental points actually follow a very smooth path with no real sign of a transition.

\begin{figure}[t]
\begin{center}
\scalebox{0.325}{\centering
\includegraphics{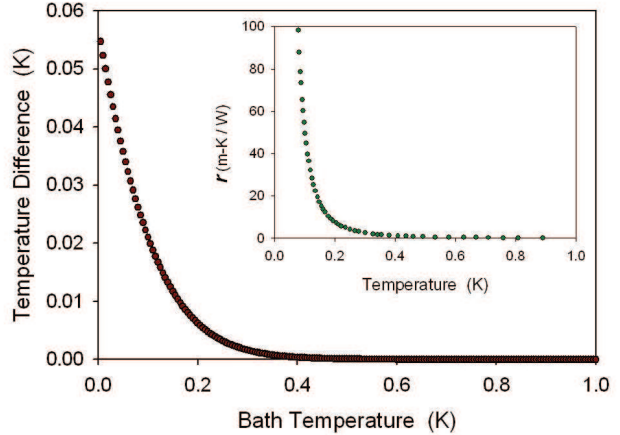}}%
\caption{Difference between junction temperature and bath temperature versus
bath temperature, as predicted by the empirical expression for $\varepsilon$.
Inset: data for thermal resistivity ($r$) of Niobium.}%
\label{Fig5}%
\end{center}
\end{figure}

As noted already, $T_{B}/\varepsilon$ represents the junction temperature
$T_{J}$, and for $\varepsilon<1$, $T_{J}>T_{B}$. Thus%
\begin{equation}
T_{J}-T_{B}=T_{B}\left[  \varepsilon^{-1}-1\right]  \label{junctiontemp}%
\end{equation}
indicates the amount by which the junction temperature is elevated above the
bath temperature. This is plotted in Fig.~\ref{Fig5}.%
Below $T_{B}\approx400mK$ the junction temperature begins to rise and, for the
experimental conditions in \cite{Yu}, would reach a maximum of about $55mK$
above the bath.

Seemingly elevated sample temperatures have been noted before
\cite{Blackburn2012}, \cite{Cristiano}, \cite{Berkley}. It is well known that
a sample subjected to input power will experience an elevation of temperature
proportional to the product of that power and the thermal resistivity of the
link to cooler surroundings. When the thermal path out of a junction becomes
more resistive, the junction temperature will rise.\ This point was made by
Kumano {\it et al.} \cite{Kumano} who noted that for samples of organic molecular
crystals characterized by extremely low thermal conductivity, a
\textquotedblleft temperature \textit{difference} (will exist) between the
sample and the thermometer\textquotedblright\ for measurements taken in a
dilution refrigerator. Similar comments concerning the liklihood of elevated
sample temperatures were expressed to us by I. Bradley \cite{Bradley}. The
inset to Fig.~\ref{Fig5} shows thermal resistivity data for Niobium
\cite{Lounasmaa}. There is a marked increase in thermal resistivity below a
few hundred millikelvins, just where our parameter $\varepsilon$ implies the
junction temperature will begin to grow significantly above $T_{B}$. Thus the
shape of the temperature elevation curve in Fig.~\ref{Fig5} strikingly echoes
the underlying thermal property of the sample material. \ 

\begin{figure}[t]
\begin{center}
\scalebox{0.3}{\centering
\includegraphics{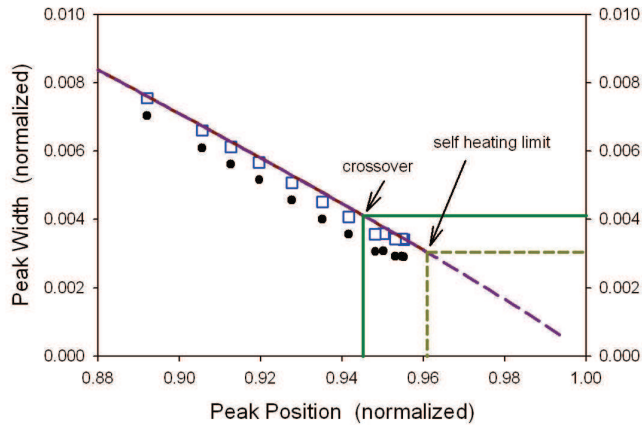}}%
\caption{Relationship between the position of the peak in the escape
probability distribution and the width of that peak (numerical simulation
using system parameters from \cite{Yu}). Dots correspond to experimental data
values from Fig.~2 in \cite{Yu}. Squares are experimental data points shifted
upwards by $1nA$. Solid lines point to the location of the crossover peak;
dashed lines point to the escape peak in the limit of zero bath temperature
(see Fig.~\ref{Fig2}) with position $0.961$ and width $0.003$.}%
\label{Fig6}%
\end{center}
\end{figure}

Finally, we comment on the relationship of SCD peaks and their widths. Voss
and Webb \cite{Voss} focused on the temperature dependence of peak width and
emphasized that the width should become independent of temperature at low
temperature because of the expected transition from TA to MQT. \ Their plot of
peak width employed a logarithmic temperature scale which of course visually
stretches the temperature axis leading to a perceptual impression of
flattening. In other words, what can seem to be evidence of width saturation
might not really be the case. \ This initial focus on peak width as opposed to
peak position has persisted in the literature. \ In truth there is no more
information to be found in the peak widths than in the peak positions. \ As
can be seen easily in Fig.~\ref{Fig1} there is a clear linkage between peak
positions and peak widths - the higher the peak position, the narrower the
peak (a similar observation was made in \cite{Wallraff}). This is shown
explicitly in Fig.~\ref{Fig6}.%

Peak positions are typically much larger numbers ($\mu A$) than widths ($nA$)
and, in addition, in the crucial domain of very low temperatures the peaks
become expecially narrow and thus subject to increased measurement error. So
the positions are more accurately known. If the width freezes, then the
position must also freeze.\ Therefore, the observed temperature dependence of
the peak positions is preferable in assessing evidence or lack of it for a
crossover from TA to MQT.

Over the years, MQT has gone from being a possibility to an apparently
confirmed theory, and experiments on Josephson junctions at millikelvin
temperatures are routinely discussed in the language of quantum mechanics. The
wells in the washboard potential are expected to possess quantized levels, and
microwaves are presumed to stimulate level transitions \cite{Blackburn2012}%
,\cite{NGJetal}; the classical thermal model is no longer considered in the
interpretation of experimental data. But we have shown here that published
data at millikelvin temperatures do agree with classical thermal activation
theory while the same experimental data are at odds with the MQT hypothesis.
As noted already, experimentally observed SCD peaks persist beyond the
crossover point shown in Fig.~\ref{Fig6}.

Our analysis indicates that classical escape models convincingly explain
observed switching current distributions in Josephson junctions at millikelvin
temperatures. We believe that this close analysis of the experimental data
reported in \cite{Yu} would apply equally to other published results for swept
bias experiments in the absense of microwaves. \ Voss and Webb \cite{Voss}
asserted that evidence for MQT should appear in SCD data from swept bias
experiments. The point is not that such evidence might be present, but that it
\textit{must} be present. However the expected unambiguous transition to
macroscopic quantum behavior is not, actually, evidenced in experimental data.
\ Of course, a more definitive test would be to perform an experiment that
reaches significantly below the supposed crossover temperature, either by
employing a sample with high $T_{cr}$ or by pushing the bath temperatures much lower.

It is worth recalling that the original analysis by Affleck \cite{Affleck81}
relied on the assumption that macroscopic quantum tunneling effects in a one
dimensional potential can be observed only if the system does not respond to
thermal effects. However, even if the Josephson excitation energies, namely
the height of the washboard potential, are above the thermal $k_{B}T$ level,
the current and phase variables fluctuate according to classical
distributions. Further, from previously published analysis
\cite{Blackburn2012} it is known that the observations of inferred quantized
energy levels inside the wells of the washboard potential at very low
temperatures can be interpreted alternatively as nonlinear resonances. Thus,
neither conditions outlined by Affleck \cite{Affleck81} for a reliable
observation of quantum effects (existence of quantized energy levels and
non-thermal statistical distributions) are convincingly present in systems of
Josephson junctions. But, as we have shown here, there is a possibility that
the observations are dominated by enhanced thermal fluctuations caused by
overheating. We believe that an important issue for future experiments will be
the determination of an effective temperature of the junctions because an
estimate of this parameter could provide information toward a possible
distinction between classical and macroscopic quantum tunnelling regimes.

This work was supported (JAB) by a grant from the Natural Sciences and
Engineering Research Council of Canada and by a grant from Wilfrid Laurier University.


\begin{thebibliography}{99}                                                                                               %


\bibitem {Tinkham}M. Tinkham, \textit{Introduction to Superconductivity}
(Dover, NY, 1996), pp. 202-205.

\bibitem {Kramers}H.A. Kramers, Physica \textbf{VII}, 284 (1940).

\bibitem {Voss}R.F. Voss and R.A. Webb, Phys. Rev. Lett. \textbf{47}, 265 (1981)

\bibitem {Blackburn2012}James A. Blackburn, Matteo Cirillo, and Niels
Gr{\o }nbech-Jensen, Phys.~Rev.~ \textbf{B 85}, 104501 (2012)

\bibitem {Yu}H.F. Yu, X.B. Zhu, Z.H. Peng, W.H. Cao, D.J. Cui, Ye Tian, G.H.
Chen, D.N. Zheng, X.N. Jing, Li Lu, S.P. Zhao, and Siyuan Han, Phys.~Rev.~
\textbf{B 81}, 144518 (2010)

\bibitem {Li}S.X. Li, W. Qiu, S. Han, Y.F. Wei, X.B. Zhu, C.Z. Gu, S.P. Zgao,
and H.B. Wang, Phys. Rev. Lett. \textbf{99}, 037002 (2007); Fig.~4

\bibitem {Cui}D.J. Cui, H.F. Yu, Z.H. Peng, W.H Cao, X.B. Zhu, Ye Tian, G.H.
Chen, D.H. Lin, C.Z. Gu, D.N. Zheng, X.N. Jing, Li Lu, and S.P. Zhao,
Supercond. Sci. Technol. \textbf{21}, 125019 (2008); Fig.~3

\bibitem {Longobardi}L. Longobardi, D. Massarotti, G. Rotoli, D. Stornaiuolo,
G. Papari, A. Kawakami, G.P. Pepe, A. Barone, and F. Tafuri, Phys.~Rev.~
\textbf{B 84}, 184504 (2011); Fig.~6

\bibitem {Yu2}H.F. Yu, X.B. Zhu, Z.H. Peng, Y. Tian, D.J. Cui, G.H. Chen, D.N.
Zheng, X.N. Jing, L. Lu, S.P. Zhao, and S. Han, Phys. Rev. Lett. \textbf{107},
067004 (2011); Fig.~3

\bibitem {Devoret}M.H. Devoret, J.M. Martinis, and J. Clarke, Phys.~Rev.~
Lett.~ \textbf{55}, 1908 (1985)

\bibitem {Q}This value of $Q$ was found to give a very good match of the MQT
peak with the TA peak at $65mK$. \ From the junction parameters given in Table
1 of \cite{Yu} and $Q=\omega_{P}RC$, we find $Q\approx18$, a value not
significantly different from our `best fit' estimate.

\bibitem {Cristiano}R. Cristiano and P. Silvestrini, Il Nuovo Cimento, Note
Brevi, \textbf{10}, 869 (1988).

\bibitem {Berkley}A.J. Berkley, H. Xu, M.A. Gubrud, R.C. Ramos, J.R. Anderson,
C.J. Lobb, and F.C. Wellstood, Phys. Rev. \textbf{B 68}, 060502 (2003); on
page 3 of this paper is the comment \textquotedblleft The $60mK$ temperature
(inferred from fitting escape-rate curves) was $40mK$ above the base
temperature, probably due to self-heating.\textquotedblright

\bibitem {Kumano}M. Kumano, Y. Ikegami, T. Sato, and S. Saito, Rev. Sci.
Instrum. \textbf{50}, 24 (1979).

\bibitem {Bradley}Ian Bradley, Department of Physics, Lancaster University -
private communication (2013).

\bibitem {Lounasmaa}O. V. Lounasmaa,\textquotedblleft Experimental Principles
and Techniques below 1K\textquotedblright, Academic Press (1974)

\bibitem {Wallraff}A. Wallraff, A. Lukashenko, C. Coqui, A. Kemp, T. Dutyand
A.V. Ustinov, Review of Scientific Instruments \textbf{74}, 3740 (2003).

\bibitem {NGJetal}J. E. Marchese, M. Cirillo, and N. Gr\o nbech-Jensen,
Phys.~Rev.~ \textbf{B74}, 174507 (2006); N. Gr\o nbech-Jensen, J. E.
Marchese, M. Cirillo, and J.A. Blackburn, Phys.~Rev.~Lett.~
\textbf{105}, 010501 (2010)

\bibitem {Affleck81}I. Affleck, Phys.~Rev.~Lett.~ \ \textbf{46}, 388 (1981).
\end{thebibliography}
\end{document}